\documentclass[12pt]{article}
\usepackage{amsmath}

\usepackage{graphics}

 \font\sevenrm=cmr7 \font\sevenbf=cmbx7
\newif\ifinexp \inexpfalse
\newcommand\I{\ifinexp \hbox{\hskip0.8pt\sevenrm i} \else
\hbox{\hskip1pt\rm i} \fi}
\newcommand\E[1]{\inexptrue \hbox{e}^{#1} \inexpfalse}
\newcommand\smallI{\hbox{\sevenrm i}}
\newcommand\smallhc{\hbox{\sevenbf C}}

\newcommand\D{\displaystyle}

\newcommand\re{\hbox{\hskip1pt\rm Re\hskip1pt}}
\newcommand\im{\hbox{\hskip1pt\rm Im\hskip1pt}}
\newcommand\hr{\hbox{\bf R}}
\newcommand\hc{\hbox{\bf C}}
\newcommand\diag{\,\hbox{\rm diag}}
\newcommand\fl{}

\newcommand\cir[1]{{\overset\circ{#1}}}
\newcommand\ee{e}
\newcommand\col{\hbox{COL}}
\newcommand\row{\hbox{ROW}}

\newtheorem{lemma}{\bf Lemma}
\newtheorem{theorem}{\bf Theorem}
\newtheorem{remark}{\bf Remark}
\newenvironment{demo}{\textbf{Proof. }}{\vskip4pt}

\begin{document}

\title{Darboux transformations and global explicit solutions
for nonlocal Davey-Stewartson I equation}

\author{Zi-Xiang Zhou\\
School of Mathematical Sciences, Fudan University,\\
Shanghai 200433, China\\
Email: zxzhou@fudan.edu.cn}

\date{}

\maketitle

\begin{abstract}
For the nonlocal Davey-Stewartson I equation, the Darboux
transformation is considered and explicit expressions of the
solutions are obtained. Like the nonlocal equations in 1+1
dimensions, many solutions may have singularities. However, by
suitable choice of parameters in the solutions of the Lax pair, it
is proved that the solutions obtained from seed solutions which are
zero and an exponential function of $t$  respectively, by a Darboux
transformation of degree $n$ are global solutions of the nonlocal
Davey-Stewartson I equation. The derived solutions are soliton
solutions when the seed solution is zero, in the sense that they are
bounded and have $n$ peaks, and ``line dark soliton'' solutions when
the seed solution is an exponential function of $t$, in the sense
that they are bounded and their norms change fast along some
straight lines.
\end{abstract}



\section{Introduction}

In \cite{bib:AbMu}, Ablowitz and Musslimani introduced the nonlocal
nonlinear Schr\"odinger equation and got its explicit solutions by
inverse scattering. Quite a lot of work were done after that for
this equation and other
equations.\cite{bib:AbMu2,bib:AbMu3,bib:Gadz,bib:HuangLing,bib:Khare,
bib:YJK,bib:LiXu,bib:Lou,bib:Zhu,bib:Sarma}

In \cite{bib:Fokashigh}, Fokas studied high dimensional equations and
introduced the nonlocal Davey-Stewartson I equation
\begin{equation}
   \begin{array}{l}
   \D\I u_t=u_{xx}+u_{yy}+2\sigma u^2\bar u^*+2uw_y,\\
   \D w_{xx}-w_{yy}=2\sigma(u\bar u^*)_{y},\\
   \end{array} \label{eq:EQ_DS}
\end{equation}
where $w$ satisfies $\bar w^*=w$. Here $\bar f(x,y,t)=f(-x,y,t)$ for
a function $f$, $*$ refers to complex conjugation. The solution of
(\ref{eq:EQ_DS}) is PT symmetric in the sense that if
$(u(x,y,t),w(x,y,t))$ is a solution of (\ref{eq:EQ_DS}), then so is
$(u^*(-x,y,-t), w^*(-x,y,-t))$. This leads to a conserved density
$u\bar u^*$, which is invariant under $x\to -x$ together with
complex conjugation.

As is known, the usual Davey-Stewartson I equation does not possess
a Darboux transformation in differential form. Instead, it has a
binary Darboux transformation in integral
form.\cite{bib:Salle,bib:MSbook} However, for the nonlocal
Davey-Stewartson I equation (\ref{eq:EQ_DS}), we can construct a
Darboux transformation in differential form. Like the nonlocal
equations in 1+1 dimensions, the solutions may have singularities.
Starting from the seed solutions which are zero and an exponential
function of $t$, we prove that the derived solutions can be globally
defined and bounded for all $(x,y,t)\in\hr^3$ if the parameters are
suitably chosen. Unlike the usual Davey-Stewartson I equation where
localized solutions are dromion solutions if the seed solution is
zero,\cite{bib:BLP,bib:Herm,bib:Hiet} the derived solutions here are
soliton solutions in the sense that there are $n$ peaks in the
solutions obtained from a Darboux transformation of degree $n$. If
the seed solution is an exponential function of $t$, the norms of
the derived solutions change a lot along some straight lines. We
call them ``line dark soliton'' solutions.

In Section~\ref{sect:LPDT} of this paper, the Lax pair for the
nonlocal Davey-Stewartson I equation is reviewed and its symmetries
are considered. Then the Darboux transformation is constructed and
the explicit expressions of the new solutions are derived. In
Section~\ref{sect:soliton} and Section~\ref{sect:darksoliton}, the
soliton solutions and ``line dark soliton'' solutions are
constructed respectively. The globalness, boundedness and the
asymptotic behaviors of those solutions are discussed.

\section{Lax pair and Darboux transformation}\label{sect:LPDT}

Consider the $2\times 2$ linear system
\begin{equation}
   \begin{array}{l}
   \D\Phi_x=\tau J\Phi_y+\tau
   P\Phi=\tau\left(\begin{array}{cc}
   1&0\\0&-1\end{array}\right)\Phi_y
   +\tau\left(\begin{array}{cc}0&-u\\v&0\end{array}\right)\Phi,\\
   \D\Phi_t=-2\I\tau^2J\Phi_{yy}-2\I\tau^2P\Phi_y+\I Q\Phi\\
   \D\qquad=-2\I\tau^2\left(\begin{array}{cc}
   1&0\\0&-1\end{array}\right)\Phi_{yy}
   -2\I\tau^2\left(\begin{array}{cc}
   0&-u\\v&0\end{array}\right)\Phi_y\\
   \D\qquad+\I\tau\left(\begin{array}{cc}
   -\tau uv-\tau w_y-w_x&u_x+\tau u_y\\
   v_x-\tau v_y&\tau uv+\tau w_y-w_x\end{array}\right)\Phi
   \end{array}\label{eq:LP}
\end{equation}
where $\tau=\pm 1$, $u,v,w$ are functions of $(x,y,t)$. The
compatibility condition $\Phi_{xt}=\Phi_{tx}$ gives the evolution
equation
\begin{equation}
   \begin{array}{l}
   \D\I u_t=u_{xx}+\tau^2u_{yy}+2\tau^2u^2v+2\tau^2uw_y,\\
   \D-\I v_t=v_{xx}+\tau^2v_{yy}+2\tau^2uv^2+2\tau^2vw_y,\\
   \D w_{xx}-\tau^2w_{yy}=2\tau^2(uv)_{y}.\\
   \end{array}\label{eq:eq}
\end{equation}

When $\tau=1$, $v=\sigma\bar u^*$ $(\sigma=\pm 1)$, (\ref{eq:eq})
becomes the nonlocal Davey-Stewartson I equation (\ref{eq:EQ_DS}).
The Lax pair (\ref{eq:eq}) becomes
\begin{equation}
   \begin{array}{l}
   \D\Phi_x=U(\partial)\Phi\overset\triangle
   =J\Phi_y+P\Phi
   =\left(\begin{array}{cc}1&0\\0&-1\end{array}\right)\Phi_y
   +\left(\begin{array}{cc}
   0&-u\\\sigma\bar u^*&0\end{array}\right)\Phi,\\
   \D\Phi_t=V(\partial)\Phi\overset\triangle
   =-2\I J\Phi_{yy}-2\I P\Phi_y+\I Q\Phi\\
   \D\qquad=-2\I\left(\begin{array}{cc}
   1&0\\0&-1\end{array}\right)\Phi_{yy}
   -2\I\left(\begin{array}{cc}
   0&-u\\\sigma\bar u^*&0\end{array}\right)\Phi_y\\
   \D\qquad+\I\left(\begin{array}{cc}
   -\sigma u\bar u^*-w_y-w_x&u_x+u_y\\
   \sigma(\bar u^*)_x-\sigma(\bar u^*)_y
   &\sigma u\bar u^*+w_y-w_x
   \end{array}\right)\Phi
   \end{array}\label{eq:LP_DS}
\end{equation}
where $\D \partial=\frac{\partial}{\partial y}$. Here $U(\partial)$
implies that $U$ is a differential operator with respect to $y$.

The coefficients in the Lax pair (\ref{eq:LP_DS}) satisfies
\begin{equation}
   \bar J^*=-KJK^{-1},\quad
   \bar P^*=-KPK^{-1},\quad
   \bar Q^*=-KQK^{-1}\label{eq:sym_coef}
\end{equation}
where $\D K=\left(\begin{array}{cc}0&\sigma\\1&0\end{array}\right)$.
Here $M^*$ refers to the complex conjugation (without transpose) of
a matrix $M$. (\ref{eq:sym_coef}) implies
\begin{equation}
   \overline{U(\partial)^*}=-KU(\partial)K^{-1},\quad
   \overline{V(\partial)^*}=KV(\partial)K^{-1}.
\end{equation}
Hence we have
\begin{lemma}\label{lemma:sym}
If $\D\Phi=\left(\begin{array}{c}\xi\\\eta\end{array}\right)$ is a
solution of (\ref{eq:LP_DS}), then so is $\D
K\bar\Phi^*=\left(\begin{array}{c}\sigma\bar\eta^*\\\bar
\xi^*\end{array}\right)$.
\end{lemma}

By Lemma~\ref{lemma:sym}, take a solution
$\D\left(\begin{array}{c}\xi\\\eta\end{array}\right)$ of
(\ref{eq:LP_DS}) and let $\D
H=\left(\begin{array}{cc}\xi&\sigma\bar\eta^*\\
\eta&\bar\xi^*\end{array}\right)$, then $G(\partial)=\partial-S$
with $S=H_yH^{-1}$ gives a Darboux
transformation.\cite{bib:ZZX,bib:GHZbook} Written explicitly,
\begin{equation}
   S=\frac 1{\xi\bar\xi^*-\sigma\eta\bar\eta^*}
   \left(\begin{array}{cc}\bar\xi^*\xi_y-\sigma\eta(\bar\eta^*)_y
   &\sigma\xi(\bar\eta^*)_y-\sigma\bar\eta^*\xi_y\\
   \bar\xi^*\eta_y-\eta(\bar\xi^*)_y&\xi(\bar\xi^*)_y-\sigma\bar\eta^*\eta_y.
   \end{array}\right).
\end{equation}
$G(\partial)$ also keeps the symmetries (\ref{eq:sym_coef})
invariant. After the action of $G(\partial)$, $(u,w)$ is transformed
to $(\widetilde u,\widetilde w)$ by
\begin{equation}
   \begin{array}{l}
   G(\partial)U(\partial)+G_x(\partial)
   =\widetilde U(\partial)G(\partial),\quad
   G(\partial)V(\partial)+G_t(\partial)
   =\widetilde V(\partial)G(\partial).
   \end{array} \label{eq:Gtransf}
\end{equation}
That is,
\begin{equation}
   \begin{array}{l}
   \D\widetilde u=u+2\sigma\frac{\bar\eta^*\xi_y-\xi(\bar\eta^*)_y}
   {\bar\xi^*\xi-\sigma\bar\eta^*\eta},\\
   \D\widetilde w=w+2\frac{(\bar\xi^*\xi-\sigma\bar\eta^*\eta)_y}
   {\bar\xi^*\xi-\sigma\bar\eta^*\eta}.
   \end{array}\label{eq:DTuv1}
\end{equation}

The Darboux transformation of degree $n$ is given by a matrix-valued
differential operator
\begin{equation}
   G(\partial)=\partial^n+G_1\partial^{n-1}+\cdots+G_n
\end{equation}
of degree $n$ which is determined by
\begin{equation}
   G(\partial)H_j=0\quad(j=1,\cdots,n)\label{eq:GH=0}
\end{equation}
for $n$ matrix solutions $H_j$ $(j=1,\cdots,n)$ of (\ref{eq:LP}). By
comparing the coefficients of $\partial^j$ in (\ref{eq:Gtransf}),
the transformation of $(P,Q)$ is
\begin{equation}
   \widetilde P=P-[J,G_1],\quad
   \widetilde Q=Q+2[J,G_2]-2[JG_1-P,G_1]+4JG_{1,y}-2nP_y.
   \label{eq:PQ}
\end{equation}

Rewrite (\ref{eq:GH=0}) as
\begin{equation}
   \partial^nH_j+G_1\partial^{n-1}H_j+\cdots+G_nH_j=0\quad
   (j=1,\cdots,n),
\end{equation}
then
\begin{equation}
   \begin{array}{l}
   \D\left(\begin{array}{cccc}G_1&G_2&\cdots&G_n\end{array}\right)
   \left(\begin{array}{cccc}\partial^{n-1}H_1&\partial^{n-1}H_2
   &\cdots&\partial^{n-1}H_n\\
   \partial^{n-2}H_1&\partial^{n-2}H_2&\cdots&\partial^{n-2}H_n\\
   \vdots&\vdots&&\vdots\\
   H_1&H_2&\cdots&H_n\end{array}\right)\\
   \D=\left(\begin{array}{cccc}-\partial^nH_1&-\partial^nH_2
   &\cdots&-\partial^nH_n\end{array}\right).
   \end{array}
\end{equation}
Write $\D H_j=\left(\begin{array}{cc}h_{11}^{(j)}&h_{12}^{(j)}\\
h_{21}^{(j)}&h_{22}^{(j)}\end{array}\right)$. By reordering the rows
and columns, we have
\begin{equation}
   \begin{array}{l}
   \D\left(\begin{array}{cccccc}(G_1)_{11}&\cdots&(G_n)_{11}
   &(G_1)_{12}&\cdots&(G_n)_{12}\\
   (G_1)_{21}&\cdots&(G_n)_{21}
   &(G_1)_{22}&\cdots&(G_n)_{22}\end{array}\right)W=-R\\
   \end{array}\label{eq:G12b}
\end{equation}
where $\D W=(W_{jk})_{1\le j,k\le 2}$, $\D R=(R_{jk})_{1\le j,k\le
2}$,
\begin{equation}
   W_{jk}=\left(\begin{array}{cccc}
   \partial^{n-1}h_{jk}^{(1)}&\partial^{n-1}h_{jk}^{(2)}
   &\cdots&\partial^{n-1}h_{jk}^{(n)}\\
   \partial^{n-2}h_{jk}^{(1)}&\partial^{n-2}h_{jk}^{(2)}
   &\cdots&\partial^{n-2}h_{jk}^{(n)}\\
   \vdots&\vdots&&\vdots\\
   h_{jk}^{(1)}&h_{jk}^{(2)}&\cdots&h_{jk}^{(n)}\\
   \end{array}\right),\label{eq:Wjk}
\end{equation}
\begin{equation}
   R_{jk}=\left(\begin{array}{cccc}
   \partial^nh_{jk}^{(1)}&\partial^nh_{jk}^{(2)}
   &\cdots&\partial^nh_{jk}^{(n)}\end{array}\right)\quad(j,k=1,2).
   \label{eq:Rjk}
\end{equation}
Solving $G$ from (\ref{eq:G12b}), we get the new solution of the
equation (\ref{eq:EQ_DS}) from (\ref{eq:PQ}). Especially,
\begin{equation}
   \widetilde u=u+2(G_1)_{12}.\label{eq:trans_u}
\end{equation}

\section{Soliton solutions}\label{sect:soliton}

\subsection{Single soliton solutions}\label{subsect:soliton1}

Let $u=0$, then $\D\Phi=\left(\begin{array}{c}
\xi\\\eta\end{array}\right)$ satisfies
\begin{equation}
   \begin{array}{l}
   \D\Phi_x=\left(\begin{array}{cc}
   1&0\\0&-1\end{array}\right)\Phi_y,\quad
   \D\Phi_t=-2\I\left(\begin{array}{cc}
   1&0\\0&-1\end{array}\right)\Phi_{yy}.
   \end{array} \label{eq:LP_DSeg}
\end{equation}
Take a special solution
\begin{equation}
   \begin{array}{l}
   \D \xi=\E{\lambda x+\lambda y-2\I\lambda^2t}
   +\E{-\lambda^*x-\lambda^*y-2\I\lambda^{*2}t},\\
   \D \eta=a\E{\lambda x-\lambda y+2\I\lambda^2t}
   +b\E{-\lambda^*x+\lambda^*y+2\I\lambda^{*2}t},
   \end{array}\label{eq:1stnxieta}
\end{equation}
where $\lambda,a,b$ are complex constants. (\ref{eq:DTuv1}) gives
the explicit solution
\begin{equation}
   \begin{array}{l}
   \D\widetilde u
   \D =\frac{2\sigma\lambda_R(a^*-b^*)}{D}
   \E{2\I\lambda_I y-4\I(\lambda_R^2-\lambda_I^2)t}
   \end{array}\label{eq:u1eg}
\end{equation}
of (\ref{eq:EQ_DS}) where
\begin{equation}
   \begin{array}{l}
   D=(2-\sigma|a|^2-\sigma|b|^2)\cosh(2\lambda_R y
   +8\lambda_R\lambda_It)\\
   \qquad+\sigma(|a|^2-|b|^2)\sinh(2\lambda_R y+8\lambda_R\lambda_It)\\
   \qquad+2(1-\sigma\re(ab^*))\cosh(2\lambda_R x)
   -2\I\sigma\im(ab^*)\sinh(2\lambda_Rx).
   \end{array}
\end{equation}
Here $z_R=\re z$ and $z_I=\im z$ for a complex number $z$. Since
$\widetilde w$ is looked as an auxiliary function in
(\ref{eq:EQ_DS}), hereafter we will study mainly the behavior of
$\widetilde u$.

Note that $\widetilde u$ is global if $|a|<1$ and $|b|<1$ since $\re
D>0$ in this case. Moreover, the peak moves in the velocity
$(v_x,v_y)=(0,-4\lambda_I)$.

\begin{remark}
The solution (\ref{eq:u1eg}) may have singularities when the
parameters are not chosen suitably, say, when $\sigma=-1$, $a=1$,
$b=-2$, or $\sigma=1$, $a=2$, $b=1/4$.
\end{remark}

Figure~1 shows a $1$ soliton solution with parameters $\sigma=-1$,
$t=20$, $\lambda=0.07-1.5\I$, $a=0.2$, $b=0.1\I$.

\begin{figure}\begin{center}
\scalebox{1.6}{\includegraphics[150,100]{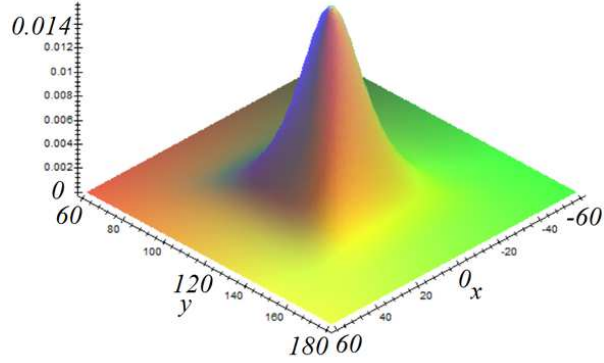}}\label{fig:stn1}
\caption{$|\widetilde u|$ of a $1$ soliton solution.}
\end{center}\end{figure}

\subsection{Multiple soliton solutions}\label{subsect:solitonn}

For an $n\times n$ matrix $M$, define $\D||M||=\sup_{x\in
\smallhc^n,||x||=1}||Mx||$ where $||\cdot||$ is the standard
Hermitian norm in $\hc^n$. The following facts hold obviously.

(i) $||MN||\le ||M||\,||N||$.

(ii) Each entry $M_{jk}$ of $M$ satisfies $|M_{jk}|\le||M||$.

(iii) $|\det M|\le||M||^n$.

(iv) If $||M||<1$, then $\D||(I+M)^{-1}||\le(1-||M||)^{-1}$.

(v) If $||M||<1$, then $\D|\det(I+M)|\ge(1-||M||)^n$.

Now we construct explicit solutions according to (\ref{eq:trans_u}).
As in (\ref{eq:1stnxieta}), take
\begin{equation}
   \begin{array}{l}
   \D\xi_k=\E{\lambda_k(x+y)-2\I\lambda_k^2t}
   +\E{-\lambda_k^*(x+y)-2\I\lambda_k^{*2}t},\\
   \D\eta_k=a_k\E{\lambda_k(x-y)+2\I\lambda_k^2t}
   +b_k\E{-\lambda_k^*(x-y)+2\I\lambda_k^{*2}t},
   \label{eq:Hjeg}
   \end{array}
\end{equation}
then $W_{jk}$'s in (\ref{eq:Wjk}) and $R_{jk}$'s in (\ref{eq:Rjk})
are
\begin{equation}\fl
   \begin{array}{l}
   (W_{11})_{jk}=\lambda_k^{n-j}\ee_{k+}
   +(-\lambda_k^*)^{n-j}\ee_{k+}^{*-1},\quad
   (W_{12})_{jk}=\sigma a_k^*(-\lambda_k^*)^{n-j}\ee_{k+}^{*-1}
   +\sigma b_k^*\lambda_k^{n-j}\ee_{k+},\\
   (W_{21})_{jk}=a_k(-\lambda_k)^{n-j}\ee_{k-}
   +b_k(\lambda_k^*)^{n-j}\ee_{k-}^{*-1},\quad
   (W_{22})_{jk}=(\lambda_k^*)^{n-j}\ee_{k-}^{*-1}
   +(-\lambda_k)^{n-j}\ee_{k-},\\
   (R_{11})_{1k}=\lambda_k^n\ee_{k+}
   +(-\lambda_k^*)^n\ee_{k+}^{*-1},\quad
   (R_{12})_{1k}=\sigma a_k^*(-\lambda_k^*)^n\ee_{k+}^{*-1}
   +\sigma b_k^*\lambda_k^n\ee_{k+},\\
   (R_{21})_{1k}=a_k(-\lambda_k)^n\ee_{k-}
   +b_k(\lambda_k^*)^n\ee_{k-}^{*-1},\quad
   (R_{22})_{1k}=(\lambda_k^*)^n\ee_{k-}^{*-1}
   +(-\lambda_k)^n\ee_{k-}\\
   (j,k=1,\cdots,n)
   \end{array}\label{eq:W22}
\end{equation}
where
\begin{equation}
   \ee_{k\pm}=\E{\lambda_k(x\pm y)\mp 2\I\lambda_k^2t}.\label{eq:ee}
\end{equation}
However, temporarily, we assume $\ee_{k+},\ee_{k-}$ $(k=1,\cdots,n)$
are arbitrary complex numbers rather that (\ref{eq:ee}) holds.

Denote $\D L=\diag((-1)^{n-1},(-1)^{n-2},\cdots,-1,1)$,
\begin{equation}
   F=(\lambda_k^{n-j})_{1\le j,k\le n},\quad
   f=(\lambda_1^n,\cdots,\lambda_n^n),
   \label{eq:V}
\end{equation}
\begin{equation}
   \begin{array}{l}
   A=\diag(a_1,\cdots,a_n),\quad
   B=\diag(b_1,\cdots,b_n),
   \end{array}\label{eq:AB}
\end{equation}
\begin{equation}
   \begin{array}{l}
   \D E_\pm=\diag(\ee_{1\pm},\cdots,\ee_{n\pm}).
   \end{array}
\end{equation}
Then
\begin{equation}
   W=\left(\begin{array}{cc}FE_++LF^*E^{*-1}_{+}
   &\sigma FB^*E_++\sigma LF^*A^*E^{*-1}_{+}\\
   LFAE_-+F^*BE^{*-1}_{-}
   &LFE_-+F^*E^{*-1}_{-}\end{array}\right),
   \label{eq:W}
\end{equation}
\begin{equation}
   R=\left(\begin{array}{cc}fE_++(-1)^nf^*E^{*-1}_{+}
   &\sigma fB^*E_++\sigma(-1)^nf^*A^*E^{*-1}_{+}\\
   (-1)^nfAE_-+f^*BE^{*-1}_{-}
   &(-1)^nfE_-+f^*E^{*-1}_{-}\end{array}\right).
   \label{eq:R}
\end{equation}
By using the identity
\begin{equation}
   \left(\begin{array}{cc}A&B\\C&D\end{array}\right)^{-1}
   =\left(\begin{array}{cc}A^{-1}+A^{-1}B\Delta^{-1}CA^{-1}&-A^{-1}B\Delta^{-1}\\
   -\Delta^{-1}CA^{-1}&\Delta^{-1}\end{array}\right)
\end{equation}
for a block matrix where $\Delta=D-CA^{-1}B$, (\ref{eq:G12b}) gives
\begin{equation}
   \Big((G_1)_{12},\cdots,(G_n)_{12}\Big)=-(RW^{-1})_{12}
   =-(R_{12}-R_{11}W_{11}^{-1}W_{12})\cir W^{-1} \label{eq:G12WR}
\end{equation}
where
\begin{equation}
   \begin{array}{l}
   \D\cir{W}=W_{22}-W_{21}W_{11}^{-1} W_{12}\\
   \D\qquad=L\Big(FE_-+LF^*E^{*-1}_{-}
   -\sigma(FAE_-+LF^*BE^{*-1}_{-})\cdot\\
   \D\qquad\quad\cdot(FE_++LF^*E^{*-1}_{+})^{-1}
   (FB^*E_++LF^*A^*E^{*-1}_{+})\Big).
   \end{array}
   \label{eq:W0}
\end{equation}

\begin{lemma}\label{lemma:lbound}
Suppose $a_j$ and $b_j$ are nonzero complex constants with
$|a_j|<1$, $|b_j|<1$ $(j=1,\cdots,n)$, $\kappa_1,\cdots,\kappa_n$
are nonzero real constants with $|\kappa_j|\ne|\kappa_k|$
$(j,k=1,\cdots,n;\,j\ne k)$, then there exist positive constants
$\delta$, $C_1$ and $C_2$, which depend on $a_j$'s, $b_j$'s and
$\kappa_j$'s, such that $|\det W|\ge C_1$ and
\begin{equation}
   \begin{array}{l}
   \D|(G_1)_{12}|\le C_2\max_{1\le k\le n}
   \frac{|\ee_{k+}|}{1+|\ee_{k+}|^2}
   \max_{1\le k\le n}\frac{|\ee_{k-}|}{1+|\ee_{k-}|^2}
   \end{array}
\end{equation}
hold whenever $|\lambda_j-\I\kappa_j|<\delta$ and $\ee_{j\pm}\in\hc$
$(j=1,\cdots,n)$.
\end{lemma}

\demo Denote $F^{-1}LF^*=I+Z$, then $Z=0$ if
$\lambda_1,\cdots,\lambda_n$ are all purely imaginary. From
(\ref{eq:W}) and (\ref{eq:W0}),
\begin{equation}
   \det W=\det(FE_++LF^*E^{*-1}_{+})\det\cir W,
   \label{eq:WWo}
\end{equation}
\begin{equation}
   \cir{W}=L(FE_-+LF^*E^{*-1}_{-})(I-\sigma\chi_-\chi_+)
   \label{eq:cirW}
\end{equation}
where
\begin{equation}
   \begin{array}{l}
   \chi_+=(FE_++LF^*E^{*-1}_{+})^{-1}(FB^*E_+
   +LF^*A^*E^{*-1}_{+})\\
   \qquad=\Xi_{1+}\Xi_{0+}^{-1}+\Xi_{0+}^{-1}(I+ZE^{*-1}_{+}
   \Xi_{0+}^{-1})^{-1}ZE^{*-1}_{+}
   (A^*-\Xi_{1+}\Xi_{0+}^{-1}),\\
   \chi_-=(FE_-+LF^*E^{*-1}_{-})^{-1}(FAE_-
   +LF^*BE^{*-1}_{-})\\
   \qquad=\Xi_{1-}\Xi_{0-}^{-1}+\Xi_{0-}^{-1}(I+ZE^{*-1}_{-}
   \Xi_{0-}^{-1})^{-1}ZE^{*-1}_{-}
   (B-\Xi_{1-}\Xi_{0-}^{-1}),\\
   \end{array}
\end{equation}
\begin{equation}
   \Xi_{0\pm}=E_\pm+E^{*-1}_{\pm},\quad
   \Xi_{1-}=AE_-+BE^{*-1}_{-},\quad
   \Xi_{1+}=B^*E_++A^*E^{*-1}_{+}.
\end{equation}

Let $\D c_0=\max_{1\le k\le n}\{|a_k|,|b_k|\}<1$. Suppose
$\D||Z||<\frac{1-c_0}2$, then we have the following estimates.
\begin{equation}
   ||A||\le c_0<1,\quad
   ||B||\le c_0<1,\label{eq:esti_AB}
\end{equation}
\begin{equation}
   ||E_{\pm}\Xi_{0\pm}^{-1}||\le 1,\quad
   ||E^{*-1}_{\pm}\Xi_{0\pm}^{-1}||\le 1,
\end{equation}
\begin{equation}
   ||\Xi_{0\pm}||\ge2,\quad
   ||\Xi_{0\pm}^{-1}||\le\frac 12,\quad
   ||\Xi_{1\pm}\Xi_{0\pm}^{-1}||\le c_0<1,
\end{equation}
\begin{equation}
   \begin{array}{l}
   \D||E^{*-1}_{+}(\Xi_{1+}\Xi_{0+}^{-1}-A^*)||
   =\max_{1\le k\le n}\frac{|a_k-b_k|\,
   |\ee_{k+}|}{1+|\ee_{k+}|^2}\le1,\\
   \D||E^{*-1}_{-}(\Xi_{1-}\Xi_{0-}^{-1}-B)||
   =\max_{1\le k\le n}\frac{|a_k-b_k|\,
   |\ee_{k-}|}{1+|\ee_{k-}|^2}\le1,
   \end{array}
\end{equation}
\begin{equation}
   ||(I+Z E^{*-1}_{\pm}\Xi_{0\pm}^{-1})^{-1}||\le||(1-||Z||\,
   ||E^{*-1}_{\pm}\Xi_{0\pm}^{-1}||)^{-1}||
   \le(1-||Z||)^{-1}\le 2.
\end{equation}
Hence $||\chi_\pm-\Xi_{1\pm}\Xi_{0\pm}^{-1}||\le||Z||$,
\begin{equation}
   ||\chi_\pm||\le c_0+||Z||\le\frac{1+c_0}2<1.
   \label{eq:esti_chi}
\end{equation}

Denote
\begin{equation}
   \pi_0=|\det F|\Big|_{\lambda_j=\smallI\kappa_j\atop{j=1,\cdots,n}},\quad
   \pi_1=||F^{-1}||\Big|_{\lambda_j=\smallI\kappa_j\atop{j=1,\cdots,n}},\quad
   \pi_2=||f||\Big|_{\lambda_j=\smallI\kappa_j\atop{j=1,\cdots,n}}.
\end{equation}
Clearly, $\pi_0$, $\pi_1$, $\pi_2$ are all positive since $\D\det
F|_{\lambda_j=\smallI\kappa_j\atop{j=1,\cdots,n}}$ is a Vandermonde
determinant. By the continuity, there exists $\delta>0$ such that
$\D\frac{\pi_0}2\le|\det F|\le 2\pi_0$, $||F^{-1}||\le2\pi_1$,
$||f||\le2\pi_2$, and $\D||F^{-1}LF^*-I||=||Z||<\frac{1-c_0}2$
whenever $|\lambda_j-\I\kappa_j|<\delta$. (\ref{eq:WWo}) and
(\ref{eq:cirW}) lead to
\begin{equation}
   \begin{array}{l}
   |\det W|\!=\!|\det F|^2\,|\det\Xi_{0+}|\,|
   \det(I+ZE^{*-1}_{+}\Xi_{0+}^{-1})|\cdot\\
   \qquad\cdot|\det\Xi_{0-}|\,|\det(I+ZE^{*-1}_{-}\Xi_{0-}^{-1})|\,
   |\det(1-\sigma\chi_-\chi_+)|\\
   \D\qquad\ge\pi_0^2(1-||Z||)^{2n}(1-||\chi_+||\,||\chi_-||)^n
   \ge\pi_0^2\Big(\frac{1+c_0}{2}\Big)^{2n}\Big(1-\Big(\frac{1+c_0}2\Big)^2\Big)^{n},
   \end{array}
\end{equation}
which is a uniform positive lower bound for any $\ee_{j\pm}\in\hc$
$(j=1,\cdots,n)$ when $|\lambda_j-\I\kappa_j|<\delta$.

By (\ref{eq:W}), (\ref{eq:R}), (\ref{eq:G12WR}) and (\ref{eq:W0}),
\begin{equation}
   \begin{array}{l}
   \D((G_1)_{12},(G_2)_{12},\cdots,(G_n)_{12})
   =-(R_{12}-R_{11}W_{11}^{-1}W_{12})\cir W^{-1}\\
   \D\qquad=-\sigma fE_+\Xi_{0+}^{-1}(I+ZE^{*-1}_{+}\Xi_{0+}^{-1})^{-1}
   (I+Z)(B^*-A^*)E^{*-1}_{+}\cir W^{-1}\\
   \qquad\quad-\sigma(-1)^nf^*E^{*-1}_{+}\Xi_{0+}^{-1}
   (I+ZE^{*-1}_{+}\Xi_{0+}^{-1})^{-1}
   (A^*-B^*)E_+\cir W^{-1}\\
   \D\qquad=-\sigma\Big(f-(-1)^nf^*+(fE_+\Xi_{0+}^{-1}+(-1)^nf^*E^{*-1}_{+}\Xi_{0+}^{-1})
   (I+ZE^{*-1}_{+}\Xi_{0+}^{-1})^{-1}Z\Big)\cdot\\
   \qquad\quad\cdot(B^*-A^*)E_+E^{*-1}_{+}\Xi_{0+}^{-1}
   (I-\sigma\chi_-\chi_+)^{-1}\Xi_{0-}^{-1}(I+ZE^{*-1}_{-}
   \Xi_{0-}^{-1})^{-1}F^{-1}L^{-1}.
   \end{array}\label{eq:G12expr}
\end{equation}
Here we have used
$I+Z=(I+ZE^{*-1}_{+}\Xi_{0+}^{-1})+ZE_+\Xi_{0+}^{-1}$. Hence, by
using (\ref{eq:esti_AB})--(\ref{eq:esti_chi}),
\begin{equation}
   \begin{array}{l}
   \D|(G_1)_{12}|\le
   8(||f||+||f^*||)||F^{-1}||\,
   ||(I-\sigma\chi_-\chi_+)^{-1}||\,||\Xi_{0+}^{-1}||\,||\Xi_{0-}^{-1}||\\
   \D\qquad\le 64\pi_1\pi_2\Big(1-\Big(\frac{1+c_0}2\Big)^2\Big)^{-1}
   \max_{1\le k\le n}\frac{|\ee_{k+}|}{1+|\ee_{k+}|^2}
   \max_{1\le k\le n}\frac{|\ee_{k-}|}{1+|\ee_{k-}|^2}.
   \end{array}
\end{equation}
The lemma is proved.

Now we consider the solutions of the nonlocal Davey-Stewartson I
equation. That is, we consider the case where $\ee_{j\pm}$'s are
taken as (\ref{eq:ee}).

\begin{theorem}
Suppose $a_j$ and $b_j$ are nonzero complex constants with
$|a_j|<1$, $|b_j|<1$ $(j=1,\cdots,n)$, $\kappa_1,\cdots,\kappa_n$
are nonzero real constants with $|\kappa_j|\ne|\kappa_k|$
$(j,k=1,\cdots,n;\,j\ne k)$, then there exists a positive constant
$\delta$ such that the following results hold for the derived
solution $\widetilde u=2(G_1)_{12}$ of the nonlocal Davey-Stewartson
I equation when $\re\lambda_j\ne 0$ and
$|\lambda_j-\I\kappa_j|<\delta$ $(j=1,\cdots,n)$.

\noindent(i) $\widetilde u$ is defined globally for $(x,y,t)\in\hr^3$.

\noindent(ii) For fixed $t$, $\widetilde u$ tends to zero
exponentially as $(x,y)\to 0$.

\noindent(iii) Let $y=\widetilde y+vt$ and keep $(x,\widetilde y)$
bounded, then $\widetilde u\to 0$ as $t\to\infty$ if
$v\ne-4\lambda_{kI}$ for all $k$.
\end{theorem}

\demo We have known that $\widetilde u$ is a solution of the
nonlocal Davey-Stewartson equation in Section~\ref{sect:LPDT}.

(i) According to Lemma~\ref{lemma:lbound}, $|\det W|$ has a uniform
positive lower bound. Hence $\widetilde u$ is defined globally.

\vskip6pt(ii) When $x\ge 0$ and $y\ge 0$,
\begin{equation}
   \begin{array}{l}
   |\ee_{k+}|\ge\E{\lambda_{kR}\sqrt{x^2+y^2}+4\lambda_{kR}\lambda_{kI}t}
   \hbox{ if }\lambda_{kR}>0,\\
   |\ee_{k+}|\le\E{-|\lambda_{kR}|\sqrt{x^2+y^2}+4\lambda_{kR}\lambda_{kI}t}
   \hbox{ if }\lambda_{kR}<0.
   \end{array}
\end{equation}
Hence $\D\max_{1\le k\le n}\frac{|\ee_{k+}|}{1+|\ee_{k+}|^2}$ tends
to zero exponentially when $x\ge 0$, $y\ge 0$ and $(x,y)\to\infty$.

Likewise, we have
\begin{equation}
   \begin{array}{l}
   |\ee_{k-}|\le\E{-\lambda_{kR}\sqrt{x^2+y^2}-4\lambda_{kR}\lambda_{kI}t}
   \hbox{ if }x\le 0,\;y\ge 0,\;\lambda_{kR}>0,\\
   |\ee_{k-}|\ge\E{|\lambda_{kR}|\sqrt{x^2+y^2}-4\lambda_{kR}\lambda_{kI}t}
   \hbox{ if }x\le 0,\;y\ge 0,\;\lambda_{kR}<0,\\
   |\ee_{k+}|\le\E{-\lambda_{kR}\sqrt{x^2+y^2}+4\lambda_{kR}\lambda_{kI}t}
   \hbox{ if }x\le 0,\;y\le 0,\;\lambda_{kR}>0,\\
   |\ee_{k+}|\ge\E{|\lambda_{kR}|\sqrt{x^2+y^2}+4\lambda_{kR}\lambda_{kI}t}
   \hbox{ if }x\le 0,\;y\le 0,\;\lambda_{kR}<0,\\
   |\ee_{k-}|\ge\E{\lambda_{kR}\sqrt{x^2+y^2}-4\lambda_{kR}\lambda_{kI}t}
   \hbox{ if }x\ge 0,\;y\le 0,\;\lambda_{kR}>0,\\
   |\ee_{k-}|\le\E{-|\lambda_{kR}|\sqrt{x^2+y^2}-4\lambda_{kR}\lambda_{kI}t}
   \hbox{ if }x\ge 0,\;y\le 0,\;\lambda_{kR}<0.
   \end{array}
\end{equation}
Lemma~\ref{lemma:lbound} implies that $\widetilde u\to 0$
exponentially as $(x,y)\to\infty$.

\vskip6pt(iii)
\begin{equation}
   |\ee_{k\pm}|=\E{\lambda_{kR}(x\pm\widetilde y)
   \pm\lambda_{kR}(v+4\lambda_{kI})t}.
\end{equation}
If $v\ne-4\lambda_{kI}$ for all $k=1,\cdots,n$, then either
$\ee_{k+}\to 0$ or $\ee_{k+}\to\infty$ for all $k=1,\cdots,n$ when
$t\to\infty$. Lemma~\ref{lemma:lbound} implies that $\widetilde u\to
0$ when $t\to\infty$. The theorem is proved.

A $3$ soliton solution is shown in Figure~2 where the parameters are
$\sigma=-1$, $t=20$, $\lambda_1=0.07-1.5\I$, $\lambda_2=0.05+2\I$,
$\lambda_3=0.1+\I$, $a_1=0.2$, $a_2=0.1\I$, $a_3=0.1$, $b_1=0.1\I$,
$b_2=-0.2$, $b_3=-0.2$. The figure of the solution appears similarly
if $\sigma$ is changed to $+1$, although it is not shown here.

\begin{figure}\begin{center}
\scalebox{1.6}{\includegraphics[150,100]{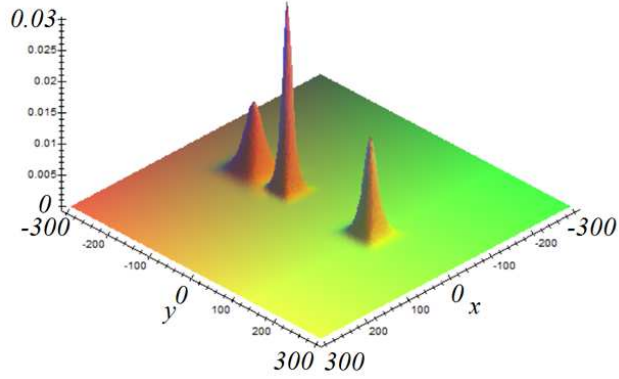}}\label{fig:stn3}
\caption{$|\widetilde u|$ of a $3$ soliton solution.}
\end{center}\end{figure}

\section{``Line dark soliton'' solutions}\label{sect:darksoliton}

\subsection{Single ``line dark soliton'' solutions}\label{subsect:darksoliton1}

Now we take
\begin{equation}
   u=\rho\E{-2\I\sigma|\rho|^2t},\quad
   w=0
\end{equation}
as a solution of (\ref{eq:EQ_DS}) where $\rho$ is a complex
constant. The Lax pair (\ref{eq:LP_DS}) has a solution
\begin{equation}
   \begin{array}{l}
   \D\left(\begin{array}{c}\E{\alpha(\lambda)x+\beta(\lambda)y+\gamma(\lambda)t}\\
   \D\frac{\lambda}{\rho}
   \E{\alpha(\lambda)x+\beta(\lambda)y
   +(\gamma(\lambda)+2\I\sigma|\rho|^2)t}\end{array}\right),
   \end{array}\label{eq:LPsndark}
\end{equation}
where
\begin{equation}
   \begin{array}{l}
   \D\alpha(\lambda)=\frac 12\Big(\frac{\sigma|\rho|^2}{\lambda}-\lambda\Big),\quad
   \beta(\lambda)=\frac 12\Big(\frac{\sigma|\rho|^2}{\lambda}+\lambda\Big),\\
   \D\gamma(\lambda)=\I(\alpha(\lambda)^2-2\alpha(\lambda)\beta(\lambda)-\beta(\lambda)^2)
   =\I\lambda^2-\frac\I 2
   \Big(\frac{\sigma|\rho|^2}{\lambda}+\lambda\Big)^2,
   \end{array}
\end{equation}
$\lambda$ is a complex constant. Note that $\alpha(-\lambda^*)=-(\alpha(\lambda))^*$,
$\beta(-\lambda^*)=-(\beta(\lambda))^*$, $\gamma(-\lambda^*)=-(\gamma(\lambda))^*$.

Now take $\D\Phi=\left(\begin{array}{c}\xi\\\eta\end{array}\right)$
where
\begin{equation}
   \begin{array}{l}
   \D \xi=\E{\alpha x+\beta y+\gamma t}+\E{-\alpha^*x-\beta^*y-\gamma^*t},\\
   \D \eta=\frac{\lambda}{\rho}\E{\alpha x+\beta y
   +(\gamma+2\I\sigma|\rho|^2)t}-\frac{\lambda^*}{\rho}
   \E{-\alpha^*x-\beta^*y-(\gamma^*-2\I\sigma|\rho|^2)t}.
   \end{array}\label{eq:darkxieta}
\end{equation}
Here $\alpha=\alpha(\lambda)$, $\beta=\beta(\lambda)$,
$\gamma=\gamma(\lambda)$. This $\Phi$ is a linear combination of the
solutions of form (\ref{eq:LPsndark}). Then (\ref{eq:DTuv1}) gives
the new solution
\begin{equation}
   \widetilde u=\rho\E{-2\I\sigma|\rho|^2t}
   \frac{\D
   \D \frac{\lambda^*}{\lambda}c_1\E{2\beta_Ry+2\gamma_Rt}
   +\frac{\lambda}{\lambda^*}c_1\E{-2\beta_Ry-2\gamma_Rt}-c_2\E{2\alpha_Rx}
   -c_2^*\E{-2\alpha_Rx}}
   {\D c_1(\E{2\beta_Ry+2\gamma_Rt}+\E{-2\beta_Ry-2\gamma_Rt})+c_2\E{2\alpha_Rx}
   +c_2^*\E{-2\alpha_Rx}}
\end{equation}
of the nonlocal Davey-Stewartson I equation where
\begin{equation}
   \begin{array}{l}
   \D c_1=1-\sigma\frac{|\lambda|^2}{|\rho|^2},\quad
   c_2=1+\sigma\frac{\lambda^2}{|\rho|^2}
   \end{array}
\end{equation}
This solution is smooth for all $(x,y,t)\in\hr^3$ if
$|\lambda|<|\rho|$.

Especially, if $\lambda$ is real, then $\gamma_R=0$, so we get a
standing wave solution.

Figure~3 shows a $1$ ``line dark soliton'' solution with parameters
$\sigma=-1$, $t=10$, $\rho=1$, $\lambda=0.3+0.1\I$. The figure on
the right describes the same solution but is upside down.

\begin{figure}\begin{center}
\scalebox{1.24}{\includegraphics[340,100]{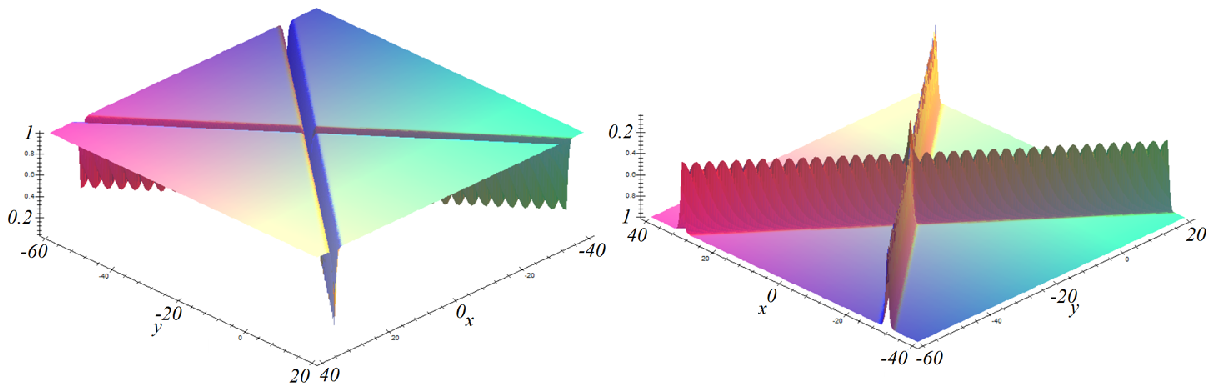}}\label{fig:1darkstn}
\caption{$|\widetilde u|$ of a $1$ ``line dark soliton'' solution.}
\end{center}\end{figure}

\subsection{Multiple ``line dark soliton'' solutions}

Now we take $n$ solutions
\begin{equation}
   \begin{array}{l}
   \D \xi_k=\E{\alpha_k x+\beta_k y+\gamma_k t}+\E{-\alpha_k^*x-\beta_k^*y-\gamma_k^*t},\\
   \D \eta_k=\frac{\lambda_k}{\rho}\E{\alpha_k x+\beta_k y
   +(\gamma_k+2\I\sigma|\rho|^2)t}-\frac{\lambda_k^*}{\rho}
   \E{-\alpha_k^*x-\beta_k^*y-(\gamma_k^*-2\I\sigma|\rho|^2)t}
   \end{array}
\end{equation}
of form (\ref{eq:darkxieta}) to get multiple ``line dark soliton''
solutions where
\begin{equation}
   \begin{array}{l}
   \D\alpha_k=\frac 12\Big(\frac{\sigma|\rho|^2}{\lambda_k}-\lambda_k\Big),\quad
   \beta_k=\frac12\Big(\frac{\sigma|\rho|^2}{\lambda_k}+\lambda_k\Big),\\
   \gamma_k=\I(\alpha_k^2-2\alpha_k\beta_k-\beta_k^2).
   \end{array}\label{eq:abc}
\end{equation}
Similar to (\ref{eq:W}) and (\ref{eq:R}), $W$ and $R$ in
(\ref{eq:G12b}) are
\begin{equation}\fl
   W=\left(\begin{array}{cc}FE_++LF^*E^{*-1}_{+}
   &-\sigma\rho^{*-1}\E{-\I\phi}(LF\Lambda E_--F^*\Lambda ^*E^{*-1}_{-})\\
   \rho^{-1}\E{\I\phi}(F\Lambda E_+-LF^*\Lambda^*E^{*-1}_{+})
   &LFE_-+F^*E^{*-1}_{-}
   \end{array}\right),\label{eq:Wdark}
\end{equation}
\begin{equation}\fl
   R=\left(\begin{array}{cc}fE_++(-1)^nf^*E^{*-1}_{+}
   &-\sigma\rho^{*-1}\E{-\I\phi}((-1)^nf\Lambda E_--f^*\Lambda ^*E^{*-1}_{-})\\
   \rho^{-1}\E{\I\phi}(f\Lambda E_+-(-1)^nf^*\Lambda^*E^{*-1}_{+})
   &(-1)^nfE_-+f^*E^{*-1}_{-}
   \end{array}\right),\label{eq:Rdark}
\end{equation}
where
\begin{equation}
   \begin{array}{l}
   E_\pm=\diag(\ee_{k\pm})_{k=1,\cdots,n},\quad
   F=(\beta_k^{n-j})_{1\le j,k\le n},\quad
   f=(\beta_1^n,\cdots,\beta_n^n),\\
   \D\alpha_k=\frac 12\Big(\frac{\sigma|\rho|^2}{\lambda_k}-\lambda_k\Big),\quad
   \beta_k=\frac12\Big(\frac{\sigma|\rho|^2}{\lambda_k}+\lambda_k\Big),\\
   \gamma_k=\I(\alpha_k^2-2\alpha_k\beta_k-\beta_k^2),
   \end{array}\label{eq:EFf}
\end{equation}
$\D \Lambda=\diag(\lambda_k)_{1\le k\le n}$,
$\phi=2\sigma|\rho|^2t$. Moreover, $\ee_{k\pm}=\E{\alpha_kx\pm
\beta_ky\pm\gamma_kt}$. However, as in the soliton case, we suppose
temporarily that $\ee_{k\pm}$'s are arbitrary complex numbers.

\begin{lemma}\label{lemma:lbounddark}
Suppose $\kappa_1,\cdots,\kappa_n$ are distinct nonzero real
numbers, then there exist positive constants $\rho_0$, $\delta$,
$C_1$ and $C_2$, which depend on $\kappa_j$'s, such that $|\det
W|\ge C_1$ and $|(G_1)_{12}|\le C_2$ hold whenever $|\rho|>\rho_0$,
$|\lambda_j-\I\kappa_j|<\delta$ and $e_{j\pm}\in\hc$
$(j=1,\cdots,n)$.
\end{lemma}

\demo Denote $F^{-1}LF^*=I+Z$, then $Z=0$ if
$\lambda_1,\cdots,\lambda_n$ are all purely imaginary. Hence $||Z||$
is small enough if
$|\lambda_1-\I\kappa_1|,\cdots,|\lambda_n-\I\kappa_n|$ are all small
enough.

Let $\D c_0=\max_{1\le k\le n}|\kappa_k|$, $\pi_3=|\det
F|\Big|_{\lambda_k=\smallI\kappa_k\atop k=1,\cdots,n}$,
$\pi_4=||F^{-1} LF||\Big|_{\lambda_k=\smallI\kappa_k\atop
k=1,\cdots,n}$. Then there exists $\delta$ with $0<\delta<c_0$ such
that $\D||Z||\le\frac12$, $\D|\det F|\ge\frac{\pi_3}2$, $||F^{-1}
LF||\le 2\pi_4$ if $|\lambda_k-\I\kappa_k|<\delta$ $(k=1,\cdots,n)$.
In this case, $|\lambda_k|<c_0+\delta<2c_0$.

From (\ref{eq:Wdark}),
\begin{equation}
   \det W=\det(FE_++LF^*E^{*-1}_{+})\det\cir W,
   \label{eq:WWodark}
\end{equation}
where
\begin{equation}
   \begin{array}{l}
   \cir{W}=LFE_-+F^*E^{*-1}_{-}
   +\sigma|\rho|^{-2}(F\Lambda E_+-LF^*\Lambda^*E^{*-1}_{+})\cdot\\
   \qquad\cdot(FE_++LF^*E^{*-1}_{+})^{-1}
   L(F\Lambda E_--LF^*\Lambda^*E^{*-1}_{-})\\
   \qquad=(1+\sigma|\rho|^{-2}F\chi_+F^{-1} LF\chi_-F^{-1} L^{-1})LF
   (I+ZE^{*-1}_{-}\Xi_{0-}^{-1})\Xi_{0-}
   \end{array}
   \label{eq:cirWdark}
\end{equation}
\begin{equation}
   \begin{array}{l}
   \chi_\pm=F^{-1}(F\Lambda E_\pm-LF^*\Lambda^*E^{*-1}_{\pm})
   (FE_\pm+LF^*E^{*-1}_{\pm})^{-1}F\\
   \qquad=\Xi_{1\pm}\Xi_{0\pm}^{-1}
   -(Z\Lambda^*+\Xi_{1\pm}\Xi_{0\pm}^{-1} Z)
   E^{*-1}_{\pm}\Xi_{0\pm}^{-1}(I+ZE^{*-1}_{\pm}\Xi_{0\pm}^{-1})^{-1},
   \end{array}\label{eq:chipm}
\end{equation}
\begin{equation}
   \Xi_{0\pm}=E_\pm+E^{*-1}_{\pm},\quad
   \Xi_{1\pm}=\Lambda E_\pm-\Lambda^*E^{*-1}_{\pm}.
\end{equation}

We have the following estimates:
\begin{equation}
   ||E_{\pm}\Xi_{0\pm}^{-1}||\le 1,\quad
   ||E^{*-1}_{\pm}\Xi_{0\pm}^{-1}||\le 1,\quad
   ||\Xi_{1\pm}\Xi_{0\pm}^{-1}||\le 2c_0,\label{eq:Echi_dark_esti}
\end{equation}
\begin{equation}
   ||\Xi_{0\pm}||\ge 2,\quad
   ||\Xi_{0\pm}^{-1}||\le\frac 12,\quad
   |\det\Xi_{0\pm}|\ge 2^n,
\end{equation}
\begin{equation}
   ||(I+Z E^{*-1}_{\pm}\Xi_{0\pm}^{-1})^{-1}||\le(1-||Z||)^{-1}
   \le 2,\quad
   |\det(I+Z
   E^{*-1}_{\pm}\Xi_{0\pm}^{-1})|\ge(1-||Z||)^n\ge\frac 1{2^n}.
\end{equation}
Hence (\ref{eq:chipm}) implies
\begin{equation}
   \begin{array}{l}
   \D||\chi_\pm||\le 2c_0+8c_0||Z||\le 6c_0,
   \end{array}
\end{equation}
\begin{equation}
   \begin{array}{l}
   \D||\chi_+F^{-1} LF\chi_-F^{-1} L^{-1}F||\le||F^{-1} LF||^2\,||\chi_+||\,||\chi_-||\le
   144c_0^2\pi_4^2.
   \end{array}\label{eq:Echi_dark_esti_end-1}
\end{equation}
By (\ref{eq:WWodark}) and (\ref{eq:cirWdark}),
\begin{equation}
   \begin{array}{l}
   |\det W|=|\det F|^2\,
   |\det\Xi_{0+}|\,|\det\Xi_{0-}|\,|\det(I+ZE^{*-1}_{+}\Xi_{0+}^{-1})|\cdot\\
   \qquad\cdot
   |\det(I+ZE^{*-1}_{-}\Xi_{0-}^{-1})|\,
   |\det(I+\sigma|\rho|^{-2}\chi_+F^{-1} LF\chi_-F^{-1} L^{-1} F)|\\
   \D\ge\frac{\pi_3^2}{4}(1-144c_0^2\pi_4^2|\rho|^{-2})>0
   \end{array}\label{eq:Echi_dark_esti_end}
\end{equation}
if $\D|\rho|>12c_0\pi_4$. Therefore, $|\det W|$ has a uniform
positive lower bound if $\D|\rho|>12c_0\pi_4$ and $|\lambda_k-\I\kappa_k|<\delta$
$(k=1,\cdots,n)$.

By (\ref{eq:G12b}), (\ref{eq:Wdark}), (\ref{eq:Rdark}),
(\ref{eq:WWodark}) and (\ref{eq:cirWdark}),
\begin{equation}
   \begin{array}{l}
   \D((G_1)_{12},(G_2)_{12},\cdots,(G_n)_{12})
   =-(R_{12}-R_{11}W_{11}^{-1}W_{12})\cir W^{-1}\\
   \D=\sigma\rho^{*-1}\E{-\I\phi}(-1)^n\Big(f\Lambda E_--(-1)^nf^*\Lambda^*E^{*-1}_{-}
   -(fE_++(-1)^nf^*E^{*-1}_{+})\cdot\\
   \D\qquad\cdot(-1)^n
   (FE_++LF^*E^{*-1}_{+})^{-1}L(F\Lambda E_--LF^*\Lambda^*E^{*-1}_{-})\Big)\cir W^{-1}\\
   \D=\sigma\rho^{*-1}\E{-\I\phi}(-1)^n
   \Big(f\Lambda E_-\Xi_{0-}^{-1}-(-1)^nf^*\Lambda^*E^{*-1}_{-}\Xi_{0-}^{-1}-\\
   \D\qquad-(-1)^n(fE_+\Xi_{0+}^{-1}+(-1)^nf^*E^{*-1}_{+}\Xi_{0+}^{-1})
   (I+ZE^{*-1}_{+}\Xi_{0+}^{-1})^{-1} F^{-1}LF\cdot\\
   \D\qquad\cdot
   (\Xi_{1-}\Xi_{0-}^{-1}-Z\Lambda^*E^{*-1}_{-}\Xi_{0-}^{-1})\Big)
   (I+ZE^{*-1}_{-}\Xi_{0-}^{-1})^{-1}F^{-1}L^{-1}\cdot\\
   \D\qquad\cdot(1+\sigma|\rho|^{-2}F\chi_+F^{-1} LF\chi_-F^{-1} L^{-1})^{-1}.
   \end{array}
\end{equation}
$(G_1)_{12}$ is bounded when $\D|\rho|>12c_0\pi_4$ and
$|\lambda_k-\I\kappa_k|<\delta$ $(k=1,\cdots,n)$ because of the
estimates (\ref{eq:Echi_dark_esti})--(\ref{eq:Echi_dark_esti_end}).
The lemma is proved.

Now we have the following theorem for the multiply ``line dark
soliton'' solution.

\begin{theorem}
Suppose $\kappa_1,\cdots,\kappa_n$ are distinct nonzero real
numbers, then there exist positive constants $\rho_0$ and $\delta$
such that the following results hold for the derived solution
$\widetilde u=u+2(G_1)_{12}$ of the nonlocal Davey-Stewartson I
equation when $|\rho|>\rho_0$ and $|\lambda_j-\I\kappa_j|<\delta$
$(j=1,\cdots,n)$.

\noindent (i) $\widetilde u$ is globally defined and bounded for
$(x,y,t)\in\hr^3$.

\noindent (ii) Suppose the real numbers $v_x,v_y$ satisfy
$\alpha_{kR}v_x\pm\beta_{kR}v_y\ne 0$ for all $k=1,\cdots,n$ where
$\alpha_k$'s and $\beta_k$'s are given by (\ref{eq:abc}), then
$\D\lim_{s\to+\infty}|\widetilde u|=|\rho|$ along the straight line
$x=x_0+v_xs$, $y=y_0+v_ys$ for arbitrary $x_0,y_0\in\hr$.
\end{theorem}

\demo (i) follows directly from Lemma~\ref{lemma:lbounddark}. Now we
prove (ii).

Since
$|\ee_{k\pm}|=\E{(\alpha_{kR}v_x\pm\beta_{kR}v_y)s
+(\alpha_{kR}x_0\pm\beta_{kR}y_0\pm\gamma_{kR}t)}$
along the straight line $x=x_0+v_xs$, $y=y_0+v_ys$,
$\alpha_{kR}v_x\pm\beta_{kR}v_y\ne 0$ implies that for each $k$,
$e_{k+}\to 0$ or $e_{k+}\to\infty$, and $e_{k-}\to 0$ or
$e_{k-}\to\infty$ as $s\to+\infty$. Let
\begin{equation}
   \begin{array}{l}
   \D \mu_k=\left\{\begin{array}{ll}\lambda_k&\hbox{ if }\alpha_{kR}v_x+\beta_{kR}v_y>0,\\
   -\lambda_k^*&\hbox{ if }\alpha_{kR}v_x+\beta_{kR}v_y<0,\end{array}\right.\\
   \D \nu_k=\left\{\begin{array}{ll}-\lambda_k&\hbox{ if }\alpha_{kR}v_x-\beta_{kR}v_y>0,\\
   \lambda_k^*&\hbox{ if }\alpha_{kR}v_x-\beta_{kR}v_y<0,\end{array}\right.\\
   \D a_k=\left\{\begin{array}{ll}\beta_k&\hbox{ if }\alpha_{kR}v_x+\beta_{kR}v_y>0,\\
   -\beta_k^*&\hbox{ if }\alpha_{kR}v_x+\beta_{kR}v_y<0,\end{array}\right.\\
   \D b_k=\left\{\begin{array}{ll}-\beta_k&\hbox{ if }\alpha_{kR}v_x-\beta_{kR}v_y>0,\\
   \beta_k^*&\hbox{ if }\alpha_{kR}v_x-\beta_{kR}v_y<0,\end{array}\right.\\
   \end{array}
\end{equation}
then
\begin{equation}
   a_k=\frac 12\Big(\frac{\sigma|\rho|^2}{\mu_k}+\mu_k\Big),\quad
   b_k=\frac
   12\Big(\frac{\sigma|\rho|^2}{\nu_k}+\nu_k\Big).\label{eq:akbk}
\end{equation}
Rewrite (\ref{eq:G12b}) as
\begin{equation}\fl
   \begin{array}{l}
   \D\left(\begin{array}{cccccc}(G_1)_{11}&\cdots&(G_n)_{11}
   &\rho^{-1}\E{\I\phi}(G_1)_{12}&\cdots&\rho^{-1}\E{\I\phi}(G_n)_{12}\\
   (G_1)_{21}&\cdots&(G_n)_{21}
   &\rho^{-1}\E{\I\phi}(G_1)_{22}&\cdots&\rho^{-1}\E{\I\phi}(G_n)_{22}\end{array}\right)
   SWS^{-1}=-RS^{-1}\\
   \end{array}\label{eq:G12bb}
\end{equation}
where $\D
S=\left(\begin{array}{cc}I_n\\&\rho\E{-\I\phi}I_n\end{array}\right)$,
$I_n$ is the $n\times n$ identity matrix.

Applying Cramer's rule to (\ref{eq:G12bb}) and using
(\ref{eq:akbk}), we have
\begin{equation}
   \lim_{s\to+\infty}\widetilde u=\lim_{s\to+\infty}(\rho\E{-\I\phi}+2(G_1)_{12})
   =\rho\E{-\I\phi}\Big(1-2\frac{\det W_1}{\det W_0}\Big)
   =\rho\E{-\I\phi}\frac{\det W_2}{\det W_0}
\end{equation}
where
\begin{equation}\fl
   W_0=\left(\begin{array}{cccccc}
   a_1^{n-1}&\cdots&a_n^{n-1}
   &\sigma|\rho|^{-2}b_1^{n-1}\nu_1&\cdots
   &\sigma|\rho|^{-2}b_n^{n-1}\nu_n\\
   a_1^{n-2}&\cdots&a_n^{n-2}
   &\sigma|\rho|^{-2}b_1^{n-2}\nu_1&\cdots
   &\sigma|\rho|^{-2}b_n^{n-2}\nu_n\\
   \vdots&&\vdots&\vdots&&\vdots\\
   a_1&\cdots&a_n
   &\sigma|\rho|^{-2}b_1\nu_1&\cdots
   &\sigma|\rho|^{-2}b_n\nu_n\\
   1&\cdots&1&\sigma|\rho|^{-2}\nu_1&\cdots&\sigma|\rho|^{-2}\nu_n\\
   a_1^{n-1}\mu_1&\cdots&a_n^{n-1}\mu_n
   &b_1^{n-1}&\cdots&b_n^{n-1}\\
   a_1^{n-2}\mu_1&\cdots&a_n^{n-2}\mu_n
   &b_1^{n-2}&\cdots&b_n^{n-2}\\
   \vdots&&\vdots&\vdots&&\vdots\\
   a_1\mu_1&\cdots&a_n\mu_n
   &b_1&\cdots&b_n\\
   \mu_1&\cdots&\mu_n&1&\cdots&1\\
   \end{array}\right),
\end{equation}
$W_1$ is obtained from $W_0$ by replacing the $(n+1)$-th row with
\begin{equation}\fl
   \left(\begin{array}{cccccc}a_1^n&\cdots&a_n^n
   &\sigma|\rho|^{-2}b_1^n\nu_1&\cdots
   &\sigma|\rho|^{-2}b_n^n\nu_n\end{array}\right),
\end{equation}
and
\begin{equation}\fl
   W_2=\left(\begin{array}{cccccc}
   a_1^{n-1}&\cdots&a_n^{n-1}
   &\sigma|\rho|^{-2}b_1^{n-1}\nu_1&\cdots
   &\sigma|\rho|^{-2}b_n^{n-1}\nu_n\\
   a_1^{n-2}&\cdots&a_n^{n-2}
   &\sigma|\rho|^{-2}b_1^{n-2}\nu_1&\cdots
   &\sigma|\rho|^{-2}b_n^{n-2}\nu_n\\
   \vdots&&\vdots&\vdots&&\vdots\\
   a_1&\cdots&a_n
   &\sigma|\rho|^{-2}b_1\nu_1&\cdots
   &\sigma|\rho|^{-2}b_n\nu_n\\
   1&\cdots&1&\sigma|\rho|^{-2}\nu_1&\cdots&\sigma|\rho|^{-2}\nu_n\\
   -\sigma|\rho|^2a_1^{n-1}\mu_1^{-1}&\cdots&-\sigma|\rho|^2a_n^{n-1}\mu_n^{-1}
   &-\sigma|\rho|^{-2}b_1^{n-1}\nu_1^2&\cdots&-\sigma|\rho|^{-2}b_n^{n-1}\nu_n^2\\
   a_1^{n-2}\mu_1&\cdots&a_n^{n-2}\mu_n
   &b_1^{n-2}&\cdots&b_n^{n-2}\\
   \vdots&&\vdots&\vdots&&\vdots\\
   a_1\mu_1&\cdots&a_n\mu_n
   &b_1&\cdots&b_n\\
   \mu_1&\cdots&\mu_n&1&\cdots&1\\
   \end{array}\right).
\end{equation}
Denote $\row_k$ and $\col_k$ to be the $k$-th row and $k$-th column
of $W_2$ respectively. The elementary transformations
\begin{equation}
   \begin{array}{l}
   \row_{n+k+1}-2\cdot\row_k\to\row_{n+k+1}\quad(k=1,\cdots,n-1),\\
   \mu_k\cdot\col_k\to\col_k\quad(k=1,\cdots,n),\\
   \sigma|\rho|^2\nu_k^{-1}\cdot\col_{n+k}\to\col_{n+k}\quad(k=1,\cdots,n),\\
   -\sigma|\rho|^{-2}\cdot\row_{n+k}\to\row_{n+k}\quad(k=1,\cdots,n),\\
   \row_k\leftrightarrow\row_{n+k}\quad(k=1,\cdots,n)
   \end{array}
\end{equation}
transform $W_2$ to $W_0$. Hence $\D\det
W_2=\prod_{k=1}^n\frac{\nu_k}{\mu_k}\det W_0$. This leads to
$\D\lim_{s\to+\infty}|\widetilde u|=|\rho|$ since
$\D\prod_{k=1}^n|\mu_k|=\prod_{k=1}^n|\nu_k|=\prod_{k=1}^n|\lambda_k|$.
The theorem is proved.

A $2$ ``line dark soliton'' solution is shown in Figure~4 where the
parameters are $\sigma=-1$, $t=10$, $\rho=1$, $\lambda_1=0.8+0.1\I$,
$\lambda_2=-0.6-0.3\I$. The figure on the right describes the same
solution but is upside down.

\begin{figure}\begin{center}
\scalebox{1.25}{\includegraphics[340,100]{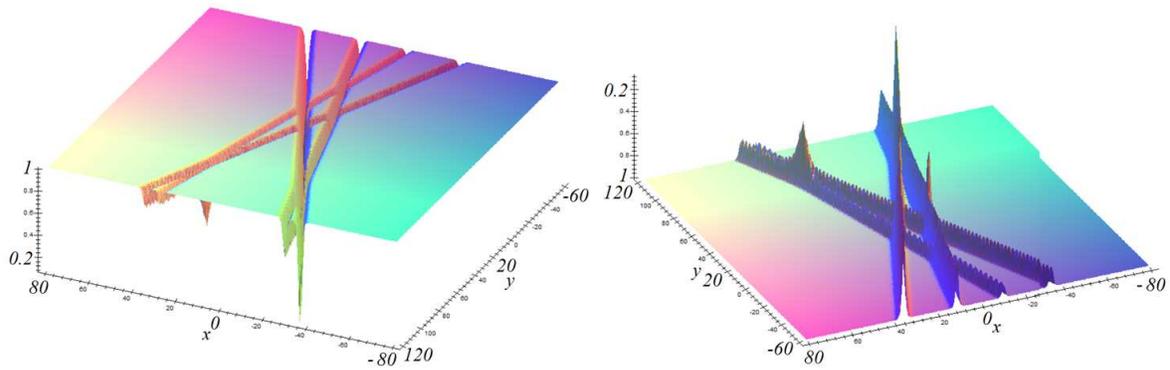}}\label{fig:2darkstn}
\caption{$|\widetilde u|$ of a $2$ ``line dark soliton'' solution.}
\end{center}\end{figure}

\section*{Acknowledgements}

This work was supported by the Natural Science Foundation of
Shanghai (No.\ 16ZR\-1402600) and the Key Laboratory of Mathematics
for Nonlinear Sciences of Ministry of Education of China. The author
is grateful to Prof.\ S.Y.Lou for helpful discussion.

\end{document}